# College Basketball: An In-depth Study of the "Foul Up 3" Dilemma

Julian Zapata-Hall, The Ohio State University

## 1. Introduction

To foul or not to foul? Within college basketball, the decision to foul or not with a 3-point lead (the *foul up 3* dilemma) significantly affects teams. For example, a single fouling decision can directly determine the outcome of a game or even an entire season. Within this study, I take a novel look at the foul up 3 dilemma, which was previously studied within a single interval of time remaining on the game clock [1, 7, 8, 9]. As opposed to adjusting in such a way, I study the foul up 3 dilemma across time. Moreover, in this study, I consider time as a two-dimensional variable (time remaining on the game clock, time remaining on the shot clock) because of the relevance of both clocks in college basketball. My research focuses on generating a coachable strategy that outlines the superior fouling decision for many different time combinations.

To achieve coachability, I sent a survey to more than 3,100 men's college basketball coaches across the country, which helped me understand the different fouling strategies and the restrictions coaches face when fouling. Based on the fouling strategies received, I divide the final 21 seconds remaining on the game clock and shot clock into 34 cells for analysis. Play-by-play (PBP) data from the 2010 to 2022 seasons of NCAA Division 1 Men's Basketball (MBB) are collected from the NCAA website to inform the study.

I calculate overtime-adjusted win percentages regarding two possible fouling decisions: foul and no foul, for each cell. The decision associated with the highest calculated overtime-adjusted win percentage is the recommended fouling decision for that cell. Additionally, I discuss the viability of a third possible fouling decision: *the lane violation* method. All code, data and detailed instructions are available at https://github.com/julianzapatahall/CBB_Up3FoulStudy to facilitate reproducibility.

The strategy I define here is coachable by any coaching staff in college basketball and is highly beneficial for the purpose of increasing winning in *up 3 situations*. The average gain associated with using the recommended strategy across all cells is 1.63 overtime-adjusted win percentage points.

### 1.1. Key Terms

Throughout this study, I use key terminology consistently to increase readability. Thus, it is critical I define the terms below.

- An *up 3 situation* occurs if the team on defense leads by exactly 3 points with fewer than 21 seconds remaining in the game.
- A *fouling opportunity* exists when the leading team is on defense and the offensive team is passing, holding, or dribbling the ball.
- A *fouling decision* is a coach's decision whether to foul or not to foul.

- An *up 3 foul* is a non-shooting common foul [3] in an up 3 situation, which follows the coach's fouling decision.

### 1.2. Notes on Terminology
- There are many types of fouls and violations in college basketball. In the context of this study, whenever I use the word foul, I am referring to an *up 3 foul*, as defined above. Additionally, when studying if a foul should be committed, I solely analyze up 3 fouls that have been committed in my data.
- Within this study, I analyze up 3 situations as defined above. Moreover, when studying which fouling decision is superior in each time cell, I only consider historical up 3 situations, from my data.

### 1.3. Calculating Overtime-Adjusted Win Percentage

In this study, I compare two distinct fouling decisions in up 3 situations: committing a foul and avoiding a foul. To compare the two, I choose a metric that is easy to understand yet statistically rigorous: overtime-adjusted win percentage. Statistician Ken Pomeroy previously used this metric [7]. It is calculated as follows:

$$OvertimeAdjustedWin\% = 100 \frac{Games\ Won}{Games\ Played},$$

where 0.5 games won are assigned to both teams in every game that extends into overtime (instead of 0 for loss or 1 for won).

### 1.4. Foul Up 3 Dilemma Overview

The foul up 3 dilemma is a problem within college basketball that concerns if a foul should be committed or avoided when leading by exactly 3 points. The context of this strategic question is as follows: The leading team is winning by 3 points but does not have possession of the ball. If the leading team can maintain its lead until there is no time remaining on the game clock, they win (since it is not the first half). If the trailing team can score a 3-point shot, they will have tied the game. To avoid this, the leading team can commit an up 3 foul, which results in the trailing team shooting a maximum of two free throws worth 1 point each [5]. Additionally, after a foul is committed, both clocks are stopped, and the shot clock is reset to its maximum value [2]. Unless the trailing team can miss their last free throw and grab the offensive rebound, the fouling team will have regained possession of the ball while maintaining the lead. Alternatively, the leading team can decide to not commit a foul, and the shot clock and game clock will continue running normally [2].

The foul up 3 dilemma is relevant across all college basketball divisions. For instance, in the championship game for the 2021-2022 NCAA Division 1 MBB season, The University of Kansas (Kansas) and The University of North Carolina at Chapel Hill (UNC) fought a close game that ended in an up 3 situation. With 19 seconds left on the game clock, UNC had possession of the ball with a 3-point deficit. Kansas decided to avoid fouling, and after two missed shots, UNC turned the ball over. Soon after, UNC regained possession with 4.3 seconds remaining on the game clock, and Kansas, once again, avoided fouling. UNC missed their last shot, and Kansas won the championship. Objectively, most teams in college basketball do not face the up 3 dilemma on such a grand scale; however, it is a grand scale dilemma. Up 3 situations test the discipline of the leading team and the resiliency of the trailing team to the highest extent. Any singular mistake by the leading team–committing a shooting foul, for example– can completely reverse the balance of the game. On the other hand, a bad offensive possession by the trailing team can eliminate any possibility of winning altogether.

## 1.5. Research goals

The first objective of this study is to develop a strategy for up 3 situations that can be coached by any coaching staff in NCAA MBB. The second objective is that the strategy recommended should maximize overtime-adjusted win percentage.

# 2. Data

To achieve coachability, I sent an anonymous survey to more than 3,100 MBB coaches across the country. Within the survey, coaches were asked to specify their fouling strategies and the variables they use to make fouling decisions. Based on the survey responses received, I divide the final seconds of a game into cells for analysis. Additionally, I choose the variables of interest: time and fouling decision. Play-by-play (PBP) data from the 2010 to 2022 seasons of NCAA Division 1 MBB were collected from the NCAA website to inform the study.

## 2.1. Survey Data

The surveys I sent helped me gain an understanding of the up 3 fouling landscape across college basketball. The surveys also asked coaches, however, about their fouling strategies in deficit situations. Each survey contained a section for coaches to mention if they had an up 3 situation strategy. Additionally, it asked for the times in which they considered fouling and the variables they weighed to generate a fouling decision.

One hundred and sixty-three total survey responses were received, and the coaches' responses were well distributed across all three MBB divisions. The most popular up 3 fouling strategy was to avoid fouling altogether. The maximum fouling time received was 20 seconds remaining on the game clock. Survey responses are available on GitHub; however, all data that could be used for personal identification have since been removed to ensure the anonymity of the responders.

## 2.2. Variable Selection

To ensure that every coaching staff can use my strategy, different restrictions on variable use are considered. For example, not all teams in college basketball have the same data accessibility. While some larger Division 1 MBB teams have access to tracking data (locations of all players and the ball at multiple frames per second), most Division 3 MBB teams do not. Additionally, pregame betting odds data, which have previously been used by Pomeroy in an up 3 study [8], are not available to most Division 3 MBB teams. Therefore, they will not be used in this study. Additionally, not all college basketball coaching staff are allowed to use electronic devices on the sideline; thus, electronic devices are not required to use the recommended strategy.

Additionally, it must be noted that communicating fouling decisions is rudimentary. It can be difficult for the coaching staff to get the attention of a player that is actively guarding the ball. Thus, during a game, most coaches resort to a hand signal that can be interpreted as a fouling decision by the team. Since there is no constant communication between the players and the coaching staff, every fouling decision must be final. Because of this, my study only contains variables that are under the full control of the coaching staff. For example, one of the most used variables in up 3 fouling studies, i.e., *free throw percentage* [8], is not under the control of the coaching staff. If the offensive team passes the ball from a player with a high free throw percentage to a player with a low free throw percentage or vice versa, there is a possibility that the recommended fouling decision would change. Thus, free throw percentage is not considered in this study.

On the other hand, I also received 163 survey responses from coaches, most of whom stated the variables they used to make a fouling decision. After reviewing the survey responses, I determined which variables are coachable during a game: time and fouling decision. Given the relevance of both clocks in college basketball, I considered time as a two-dimensional variable: (time remaining on game clock, time remaining on shot clock).

These two variables are the independent variables of my study. The target variable is an overtime-adjusted win percentage since it is interpretable by any college basketball coach. Because it measures a chance at winning, its relevance is clear.

### 2.3. Focus: Final 21 Seconds

When selecting the focus area of my study, I rely heavily on the coach survey responses. The narrowest game clock time interval associated with fouling was 2 seconds, and many coaches felt comfortable with this interval. Additionally, many coaches avoid fouling with 3 seconds or fewer on the shot clock since there is an increased chance of committing a shooting foul. Thus, I divide the game clock and shot clock values into 3-second cells. On the other hand, coaches considered fouling in the final 20 seconds remaining on the game clock; I aim to study a wider interval. Within this study, I focus on the final 21 seconds remaining on the game clock since it is divisible into 3-second cells.

The following 34-cell design was chosen, and the numbers within the cells are merely cell identifying numbers ranging from 1 to 34.

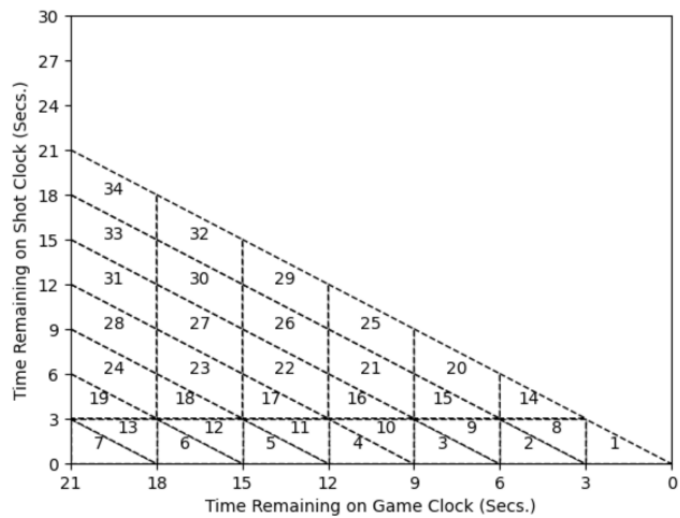

**Figure 1.** Graphical representation of the final seconds of a college basketball (CBB) game.

Figure 1 is the grid graph, which is the focal point of my study. Firstly, it is a graphical representation of the final 21 seconds of college basketball games. Secondly, the grid graphs pattern emulates the two distinct ways in which time moves within the final seconds of a college basketball game; see Figure 2 and Figure 3 below. Thirdly, the grid graph respects the restrictions of all college basketball coaching staff.

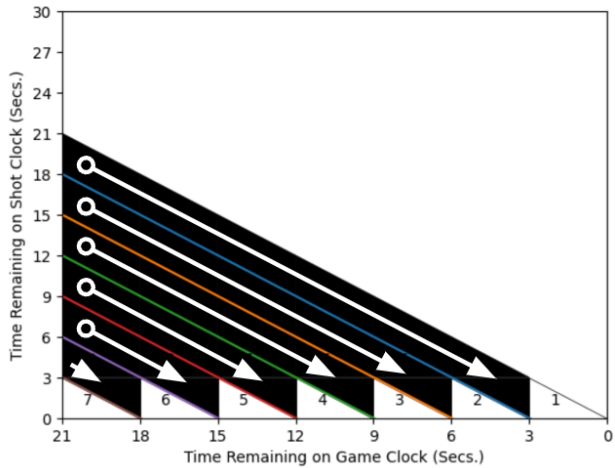

**Figure 2.** Joint behavior of the shot clock and game clock within the final seconds of a CBB game.

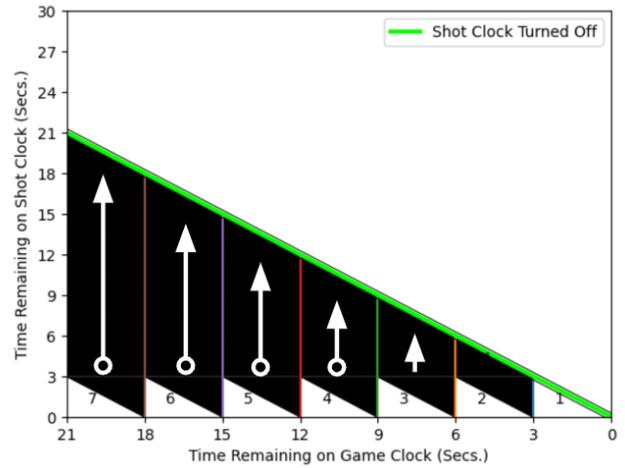

**Figure 3.** Behavior of the shot clock when a foul occurs within the final seconds of a CBB game

Within this study, I find the up 3 fouling decision that maximizes the overtime-adjusted win percentage for each of the 34 cells defined in the graph. When filled, the grid graph is green for cells in which committing a foul maximizes the overtime-adjusted win percentage and red for cells in which avoiding a foul maximizes overtime-adjusted win percentage. The cells to be colored are cells 1 through 13. Since there are fewer than 3 seconds on the shot clock in each of the cells, there is a high risk of committing a shooting foul. Thus, these cells are colored red as shown below in Figure 4.

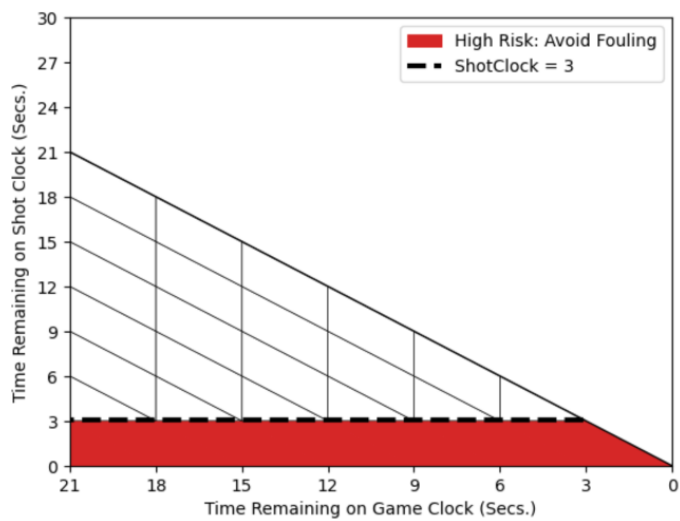

**Figure 4.** With less than 3 seconds left on the shot clock, fouls are likely to be deemed a shooting foul.

## 2.4. Play-By-Play Data Survey Data

To perform the analyses of interest, I use 13 seasons (2009-2010 to 2021-2022) of play-by-play (PBP) data from NCAA Division 1 MBB. The data were scraped directly from the NCAA website using Python to create a PBP dataset. Within the dataset, there are over 58,000 games and over 9 million plays. In short, the dataset contains all the plays that are relevant for calculating basic basketball statistics. For example, every shot attempt–made or missed–is recorded, as is every foul committed by both teams.

Within this study, I use the PBP dataset to calculate the overtime-adjusted win percentage statistic associated with each fouling decision of interest and generate an up 3 fouling strategy. The PBP dataset's size allows me to exclude human error when generating a strategy. A coach, who has experienced dozens of up 3 fouls, is biased through his or her own experiences when making up 3 fouling decisions. On the other hand, the PBP dataset contains more than 1,000 up 3 fouls, which I use to inform my analyses.

# 3. Methodology

In this study I color the grid graph, displaying the best fouling decision for each of the 34 cells. Thirteen of these cells have already been colored red; I use a statistical approach to color the remaining 21 cells.

## 3.1. Initial Modifications to the PBP Dataset

1. The dataset is duplicated by game and team such that every row of data (basketball play) appears twice in the dataset: once from the point of view of the home team and once from the away team's point of view. To differentiate each of these points of view, a column of team/game identifiers is created.
2. A column of points is added to the dataset. These points show the time immediately before the play occurs.
3. A column of line segments is added to the dataset. These line segments join the starting point with the ending point of the previous play.
4. A fouling opportunity column is created in the dataset; see Section 1.1. This column displays if there is a fouling opportunity in the current play.

## 3.2. Studying Each Cell

Fouling decisions are compared for each of the remaining 21 cells. Fixing one of the cells, following steps are taken to compare fouling decisions within this cell.

### 3.2.1. Up 3 Situation Cell-Based Modifications

5. A new column is added that contains the cell of interest for each row (play) within the PBP dataset and is used to perform efficient computations.
6. A column of intersections is calculated that contains the intersection between the line segment and cell columns. Effectively, if the intersection value for a specific play is non-empty, then, during some point in the previous play, the game clock and shot clock values lied within the cell. If the intersection value is empty, the opposite is true.
7. A second column of intersections is calculated that contains the intersection between the point and cell columns. Effectively, if the intersection value of a specific play is non-empty, then the previous play must have ended within the cell. If the intersection value is empty, the opposite is true.
8. By using the intersection and fouling opportunity columns, it is determined if one of the two fouling decisions was followed. The team/game identifiers that follow each of the fouling decisions are stored separately.

### 3.2.2. Calculating Overtime-Adjusted Win Percentages

9. By using the team/game identifiers, an overtime-adjusted win percentage is calculated for the avoid fouling decision. For committing a foul, a single overtime-adjusted win percentage value is calculated for the cell column. All cells within the column share the same valu. Overtime-adjusted win percentages are calculated as defined in Section 1.3 for each fouling decision. These two values are compared, and the cell is colored green if fouling is the superior decision or red if avoiding to foul is the best decision.

These nine steps are repeated until all 34 cells have received a recommended fouling decision.

## 4. Elements of the Foul Up 3 Dilemma

To better illustrate the foul up 3 dilemma I discuss its different elements separately and detail the distribution of up 3 fouling across the 13 different seasons in the PBP dataset. Next, I present two graphs, which show the overtime-adjusted win percentage values for the two fouling decisions in each of the 34 cells. Lastly, I discuss an alternative fouling method in detail, the lane violation approach.

### 4.1. Foul Distribution

Fouls are well distributed amongst the teams in the PBP dataset. Between 2010 and 2022, 1644 up 3 fouls were committed by 339 different teams. The team which committed the most fouls was: University of Richmond with 20.

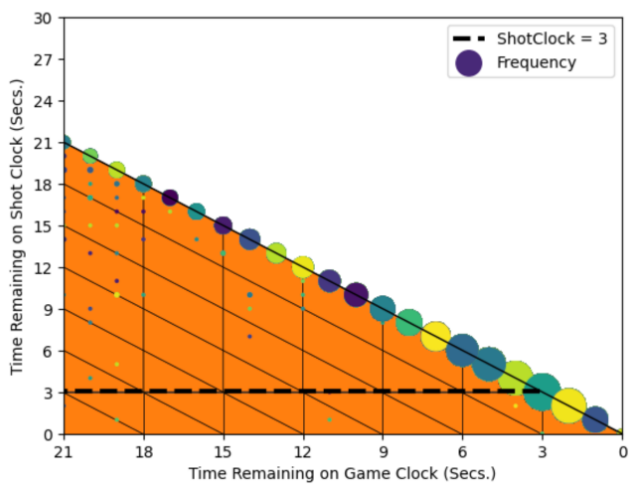

**Figure 5.** Scatter plot for up 3 fouls from the 2010-2022 CBB seasons, scaled by frequency.

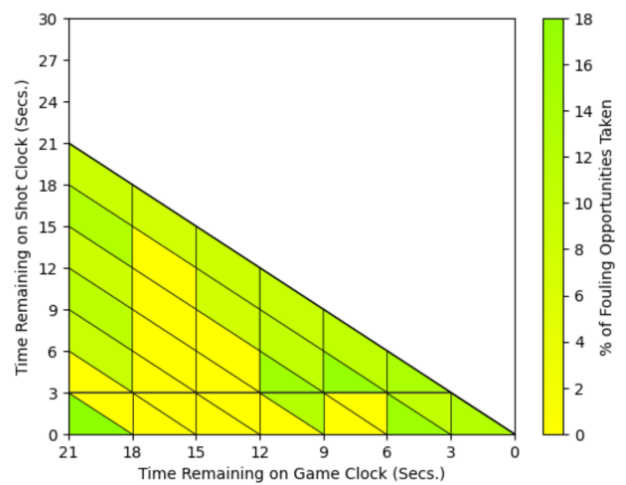

**Figure 6.** Percentage of fouling opportunities taken in the final seconds of a CBB game.

Figure 5 shows fouls committed when leading by 3 points. Most up 3 fouls occur when the shot clock is turned off; see Figure 3. The second largest group of up 3 fouls lies in the first cell column. On the other hand, few fouls were committed in any of the other cells. In many cells no fouls were executed. This, however, is not an obstacle for calculating overtime-adjusted win percentages; see Section 4.2 below. Lastly, some fouls were committed with fewer than 3 seconds remaining on the shot clock. I consider these fouls strategic mistakes and exclude them from my analyses; see Section 1.1.

Figure 6 displays the percentage of fouling opportunities taken within each cell. For example, in a hypothetical cell with 300 fouls committed in 3,000 fouling opportunities, the percentage of fouling opportunities taken would be 10%. Values within Figure 6 range from 0% to 18%. Thus, avoiding a foul is the most popular fouling decision across all cells. Additionally, Figure 6 shows that with fewer than 3 seconds remaining on the shot clock teams avoid fouling. Near the end of the game, however, teams tend to commit more of these fouls. This is likely because of an increased level of stress.

## 4.2. Choosing to Foul

The first of the two fouling decisions in my study is committing a foul. Every coach who completed my survey specified that they commit strategic fouls in some situations within a game. Thus, they have a methodology in place that could be used to perform fouls in up 3 situations.

It is insightful to note what a coach stands to gain from committing an up 3 foul at different moments near the end of the game. In this section, I analyze the overtime-adjusted win percentage associated with fouling in each of the cells of interest.

First, however, I provide an example on how to interpret overtime-adjusted win percentages associated with fouling. For example, the correct interpretation of an overtime-adjusted win percentage of 85% in Cell A is as follows:
The overtime-adjusted win percentage across all teams that committed a foul in Cell A within the PBP dataset is 85%.

For situations like those considered in this study, a reasonable interpretation for an overtime-adjusted win percentage of 85% in Cell A is:
If a team commits a foul in Cell A, at that moment they will have an 85% chance of winning.

Figure 7 shows that overtime-adjusted win percentages behave in a particular way when a foul is committed within the final seconds of a college basketball game. If a team commits a foul with a fixed number of seconds left on the game clock, their overtime-adjusted win percentage will be the same across all possible shot clock values. This is because when a foul is committed within the last 30 seconds remaining on the game clock the shot clock is turned off (the shot clock value is reset to its maximum value) [2]. Therefore, every foul committed at a given game clock value is equivalent in basketball outcomes. See Figure 3 for a detailed view into the behavior of the shot clock when a foul is committed. Thus, I assign the overtime-adjusted win percentage value for each cell column to each of its cells.

These overtime-adjusted win percentages associated with fouling range from 87% to 93% across all cells (Figure 7).

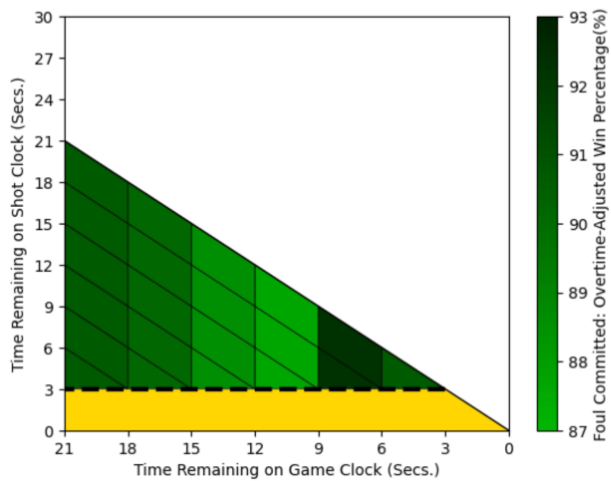

**Figure 7.** Overtime-adjusted win percentages for a team that commits a foul in the final seconds of a CBB game.

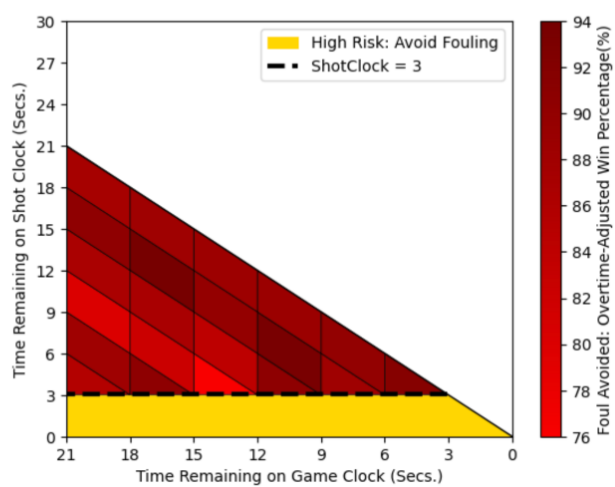

**Figure 8.** Overtime-adjusted win percentages for a team that avoids fouling in the final seconds of a CBB game.

### 4.3. Choosing Not to Foul

The second of the two fouling decisions in my study is avoiding to foul. In each of the cells defined, not fouling is the most popular decision amongst coaches; see Figure 6. In this section, I analyze the overtime-adjusted win percentages that result from avoiding to foul in each cell.

First, however, I provide an example on how to interpret overtime-adjusted win percentages associated with avoiding to foul. For example, the correct interpretation for an overtime-adjusted win percentage of 85% in Cell A is as follows:
The overtime-adjusted win percentage across all teams that avoided committing a foul in Cell A within the PBP dataset is 85%.

For situations like those considered in this study, a reasonable interpretation for an overtime-adjusted win percentage of 85% in Cell A is:
If a team avoids committing a foul in Cell A, at that moment they will have an 85% chance of winning.

The overtime-adjusted win percentages associated with not fouling in the cells of interest range from 76% to 94% (Figure 8). Two main patterns can be observed in this Figure 8. First, as the time remaining on the shot clock decreases, the win percentage associated with not fouling decreases. On the other hand, within each of the diagonal rows of cells, the win percentage remains approximately constant. Read Section 4.5 for a detailed discussion of this trend.

### 4.4. The Lane Violation Approach

There exists an alternative to the two fouling decisions in this study: the lane violation approach, which was popularized by college basketball coach Bob Walsh. It is essentially a more effective method of committing a foul in an up 3 situation. The lane violation approach, however, is highly risky since many coaches and referees believe that it exploits the current NCAA MBB rulebook. Bob

Walsh himself, has been threatened with receiving a technical foul for applying this strategy during a game, which is very detrimental to the leading team in an up 3 situation. A technical foul results in the trailing team shooting one or two free throws and receiving possession of the ball [6]. It is hard to analyze the benefits of the lane violation method since there exists a real risk of getting a technical foul. Thus, I discuss the method in a general sense but do not include it in any analyses.

The lane violation approach starts with a regular foul. Once the trailing team is going to attempt their final free throw, the leading team abandons their reglementary free throw positions and floods the paint. This is called a lane violation. According to the current NCAA MBB rulebook, this violation forces the trailing team to shoot another free throw if the previous one was missed [5]. The lane violation method consists of performing a lane violation until the free throw is made. Essentially, the lane violation method eliminates the possibility of an offensive rebound on the last free throw attempt and guarantees that the fouling team will regain the possession of the ball while maintaining their lead.

If not for the risk of a technical, the method would be a useful fouling approach. In previous analyses, I observed that the superior approach with 6 to 9 seconds remaining on the game clock is the lane violation method. In most other cases, however, regular fouling is better than performing the lane violation method.

### 4.5. Correlation Between Cells

Note that in this section, a perspective identifies a team/game combination. In each college basketball game, two perspectives exist, one for each team.

Each of the 34 cells I define is studied separately. The overtime-adjusted win percentage calculations I perform do not consider any adjacent cells or any other cells. Every game is counted at most once per cell. Thus, the outcomes within each cell are independent from one another. Since I count every relevant perspective once per cell, however, *chronologically adjacent cells* are highly correlated in outcomes. For example, if a foul was avoided in cell 29, that perspective will pass to cell 25; see Figures 1 and 2. If the perspective resulted in a win, then the win will be counted in both cells 29 and 25. This is the reason behind the strong correlation in each of the diagonal rows of Figure 8. In this study, however, all cells are studied separately to find the best decision within each cell. Thus, the outcome dependence between chronologically adjacent cells does not pose an issue.

## 5. Results

### 5.1. Recommended Strategy

The overtime-adjusted win percentages for the two fouling decisions are compared in each cell. In every cell, I select the superior fouling decision. This process is repeated until every cell is colored green (a foul should be committed) or red (a foul should be avoided). Note that I treat all recommended fouling decisions equally. A table showing all overtime-adjusted win percentages and sample sizes is available on GitHub.

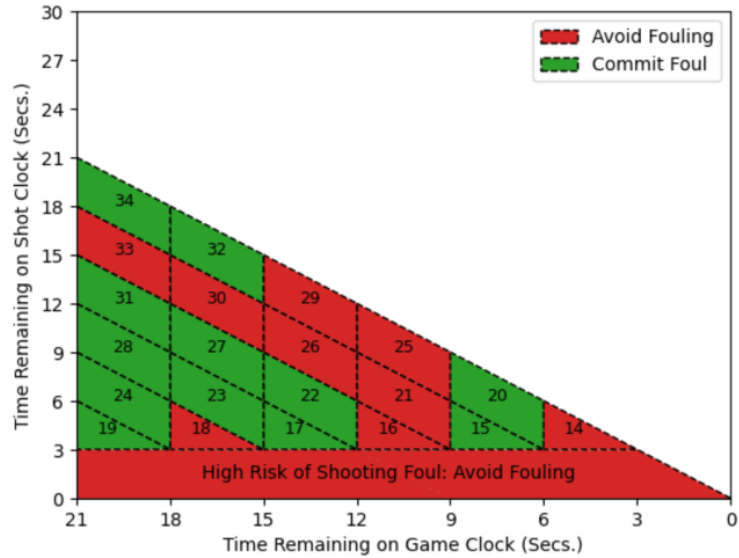

**Figure 9.** Recommended up 3 fouling strategy for the final seconds of a CBB game.

Figure 9 is my recommended strategy. The numbers within the cells are merely cell identifying numbers ranging from 14 to 34; see Figure 1. Firstly, it must be noted that the final 12 seconds remaining on the game clock match previous conceptions of when a foul should be committed. Interestingly, however, there are more green cells as the time left on the game clock increases. This suggests that fouling earlier may be beneficial. Additionally, when considering this graph alongside the two fouling decision graphs, a pattern emerges. Figure 7 shows that the overtime-adjusted win percentages associated with fouling remain constant between 87% to 93%. Figure 8 shows that the overtime-adjusted win percentages associated with not fouling dip under 80% as the shot clock decreases. Thus, fouling in the cells with a low shot clock (e.g., cell 22 and cell 27), results in the highest competitive benefit.

### 5.2. Strategy Usage
With the goal of leveraging the most data possible, I used 13 seasons of Division 1 PBP data for my analyses. Coaches across all NCAA MBB divisions stated they were comfortable using a strategy developed from this dataset.

How does a coach use my strategy?
My recommended strategy can be used within a game or practice session. It is essential, however, that a coach understands how time behaves within the grid graph before using the strategy during a game. To use the strategy, a coach must simply have a copy of Figure 9 readily available. During an up 3 situation, the coach should look at the graph and recognize the current time cell. If the cell is green, he or she should communicate that a foul should be committed. If the cell is red, he or she should communicate that a foul should not be committed.

### 5.3. Impact
My recommended strategy adds structure to the foul up 3 dilemma and provides a guideline every coach can follow in a high-pressure environment. It is reliable and founded. Moreover, a coach can only recall a limited amount of up 3 situations when making a fouling decision. This study allows coaches to access up 3 situations in 13 seasons of Division 1 MBB, thus improving their ability to make decisions in up 3 situations.

To measure the impact of my recommended strategy, I calculated the overtime-adjusted win percentage point gain from following my recommended decision in each cell. For each cell, I subtracted the average overtime-adjusted win percentage across both decisions from the overtime-adjusted win percentage of my recommended decision. By averaging gain across cells, I find that if a coach follows my recommended decision in each cell, he or she gains 1.63 overtime-adjusted win percentage points on average.

## 6. Summary and Future Work
This research not only creates an actionable fouling strategy for all NCAA MBB coaching staff but also introduces groundbreaking methodology that allows for the comparison of thousands of different strategic approaches. While these results are specific to current NCAA Men's Basketball, the methodology can be applied to generate an NBA, WNBA and NCAA Women's Basketball fouling strategy. The methodology can also be extended to football and address the fourth-down and two-point conversion problems in the NFL and college football, thereby elevating strategy in both sports.

## 7. Acknowledgements
This study would not be what it is without the support of these great professionals:
- Bob Walsh (college coach, St John's University Red Storm)
- Chris Murphy (college coach, Union College Garnet Chargers)
- Dr. Laurie Shaw (data scientist, Manchester City)
- Dr. Roger Hoerl (statistician, Union College)
- Dr. Sebastian Kurtek (statistician, The Ohio State University)
- Dr. Wayne Winston (statistician, Indiana University)
- Ken Pomeroy (statistician, kenpom.com)
- Kevin Weckworth (college coach, Hartford Hawks)
- Sudarshan Gopaladesikan (data scientist, Atalanta BC)
- 163 NCAA Men's Basketball Coaches / Anonymous Survey Responses.

## References

[1] Ezekowitz, J. (2010, August 24). Up Three, Time Running out, Do We Foul? The First Comprehensive CBB Analysis. *The Harvard Sports Analysis Collective*. https://harvardsportsanalysis.wordpress.com/2010/08/24/intentionally-fouling-up-3-points-the-first-comprehensive-cbb-analysis/.

[2] NCAA Publications. (2023a). *2023-24 Men's Rules Book*. NCAA Publications. https://www.ncaapublications.com/productdownloads/BK24.pdf, 36-39.

[3] NCAA Publications. (2023b). *2023-24 Men's Rules Book*. NCAA Publications. https://www.ncaapublications.com/productdownloads/BK24.pdf, 48.

[4] NCAA Publications. (2023c). *2023-24 Men's Rules Book*. NCAA Publications. https://www.ncaapublications.com/productdownloads/BK24.pdf, 84-85.

[5] NCAA Publications. (2023d). *2023-24 Men's Rules Book*. NCAA Publications. https://www.ncaapublications.com/productdownloads/BK24.pdf, 95-96.

[6] NCAA Publications. (2023d). *2023-24 Men's Rules Book*. NCAA Publications. https://www.ncaapublications.com/productdownloads/BK24.pdf, 97-102.

[7] Pomeroy, K. (2013, February 12). Yet Another Study about Fouling When Up 3. *The Kenpom.com Blog*. https://kenpom.com/blog/yet-another-study-about-fouling-when-up-3/

[8] Pomeroy, K., & Safir, J. (2020, July 21). The guide to fouling when leading (or tied). *The Kenpom.com Blog.* https://kenpom.com/blog/the-guide-to-fouling-when-leading-or-tied/

[9] Winston, W. L. (2012). Mathletics: How gamblers, managers, and sports enthusiasts use mathematics in baseball, basketball, and football. Princeton, NJ: Princeton University Press, 631-636.